\documentclass[a4paper,20pt]{article}
    \usepackage[T1]{fontenc}
    \usepackage{graphicx}
    \usepackage{epsfig}
    \usepackage{amsmath}
    \usepackage{amsfonts}
    \renewcommand{\abstract}{}
\begin{document}
\makeatletter
\renewcommand{\@oddhead}{\textit{YSC'14 Proceedings of Contributed Papers} \hfil \textit{E.Yu. Bannikova, M.S. Mykhailova}}
\renewcommand{\@evenfoot}{\hfil \thepage \hfil}
\renewcommand{\@oddfoot}{\hfil \thepage \hfil}
\fontsize{11}{11} \selectfont

\title{Distribution of X-ray Emission from Jet Knots of 3C273}
\author{\textsl{E.Yu. Bannikova$^{1, 2}$, M.S. Mykhailova$^{2}$}}
\date{}
\maketitle
\begin{center} {\small $^{1}$Radio Astronomy Institute NAS of Ukraine \\
$^{2}$Karazin Kharkov National University \\ bannikova@astron.kharkov.ua,
aniramtiger@rambler.ru}
\end{center}

\begin{abstract}
The jet of the quasar 3C273 is observed at different spectral
bands. This jet has a knot structure. Jet radiation in radio and
optical bands are connected with synchrotron mechanism, while the
emission mechanism producing the X-rays is controversial. We
suppose that the X-rays observed for two knots nearest to the
quasar  can originate from the inverse Compton scattering of
external source radiation on relativistic electrons. But in the
jet region with constant low X-ray intensity the inverse Compton
scattering on cosmic microwave background photons (IC/CMB) is
essential because the energy density of external source decreases.
Upon this scenario the constraints on the angle between the jet
axis and line sight values have been obtained. Also, some physical
parameters for two nearest knots of the jet of 3C273 have been
estimated.
\end{abstract}

\section*{Introduction}
\indent \indent The jet of 3C273 was observed with high resolution
by MERLIN \cite{merlin}, by the Hubble space telescope \cite{HSTM}
and Chandra \cite{Sambruna} at the radio, optical and X-ray bands
correspondingly. Through this work $H_0=75$ km/s/Mpc, so $1''$
corresponds to $2.37$ kpc at the distance of 3C273 \cite{HSTM}.

\section*{Jet morphology}
\indent \indent In the radio band the jet indicates a 'cocoon
structure'. The intensity in the jet increases in the direction away
from 3C273. In the optical band $10''$-jet is most narrow. Within
this jet knots that are location shocks occur where electrons are
accelerated at relativistic speeds. Shocks can arise by
Kelvin-Helmholtz instability at the boundary between jet matter and
ambient. These knots have almost identical brightness and size
$0.1''-0.5''$. The first knot is detected at the distance $13''$
from the core. To indicate the knots we used notation introduced
into practice by Bahcall et al. \cite{HSTM}. The knots A3, In1, In2,
Ex1 might not belong to the jet because they are not detected in the
radio and X-ray jet images \cite{HSTM}. In the X-ray band the jet
has length of $8''$. Only two optical knots can be found in it. The
remaining jet region has low constant X-ray intensity (see Fig.1).
\begin{figure}
  \centering
  \includegraphics[width=8cm,height=8cm]{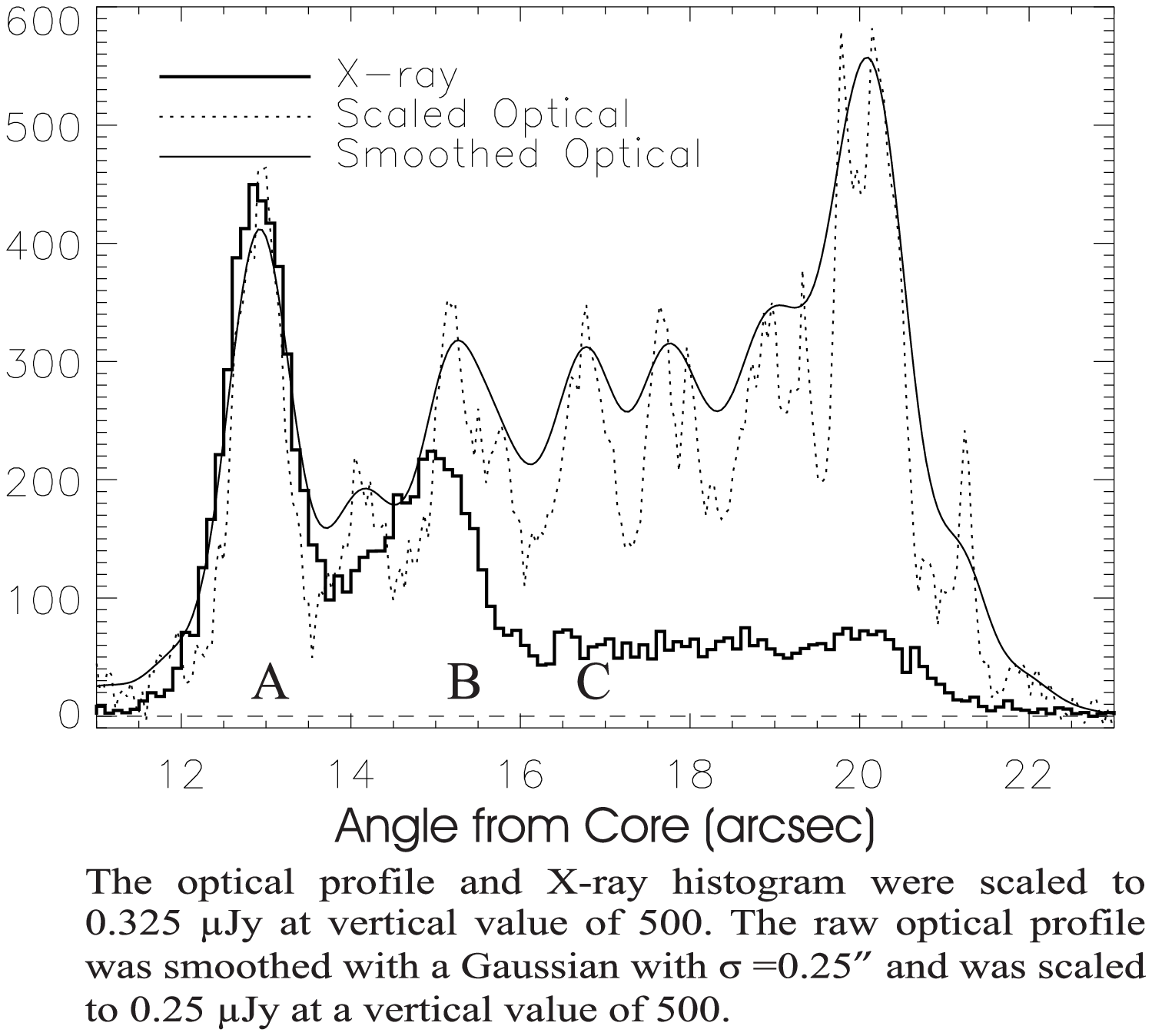}\\
  \caption{Distribution of the X-ray and optical emission along the axis of the jet of
  3C273
  \cite{Marshall}.}
  \label{3C273}
\end{figure}

\section*{Suggested idea}
\indent \indent The possibilities for the origin of the X-rays might
be \cite{Sambruna,Harris}: 1) synchrotron emission from a second,
more energetic population of electron; 2) inverse Compton scattering
(IC) of photons external to the jet; 3) IC of radio-optical
synchrotron photons. These theories cannot explain why the ratio of
X-ray to radio observed intensities decreases with distance from
3C273. But this fact can be important for study of the jet
conditions. And so, in our work it is mostly crucial.

The most plausible mechanism of X-ray emission is IC. The observed
intensity distribution was taken from work \cite{Marshall}. It
shows that two X-ray intensity peaks correspond to the nearest to
the quasar optical knots, A1 and B1. Then this intensity decreases
to a lower level but it is still detectable until it reaches the
terminating optical knot. While in the optical band the knots have
almost identical intensities. Thus, the idea of explaining the
X-ray intensity distribution observed along the jet of 3C273 is
the following \cite{Bannikova}. In different jet regions
scattering photons have different origin. For the nearest to the
quasar jet region there is a heterogeneous radiation field. While
for the remaining jet the homogeneous radiation field
predominates. In this case, X-ray emission of the two nearest
knots originates from IC of quasar photons. For the remaining
knots X-rays arise from the same scattering of cosmic microwave
background photons. Thus, the basic term was formulated
(\ref{eq1}):
\begin{equation}\label{eq1}
\begin{cases}
    W_{ES} > W_{CMB}, & \text{$r_{arc} < 16''$}, \\
    W_{ES} < W_{CMB},& \text{$r_{arc} > 16''$},
\end{cases}
\end{equation}
where $W_{ES}$ is the energy density of quasar emission, $W_{CMB}$
is the energy density of cosmic microwave background, $r_{arc}$ is
the project distance from the quasar in arcseconds, $16''$ was taken
because after this distance from 3C273 the X-ray emission have
almost constant value (see Fig.1).

Using the energy densities, the integral values can be considered.
The observed intensity decline from the knot A1 to knot B1 might be
due to the fact that the ratio of electron concentration in these
knots is $N_A / N_B = 2$. Electron concentration in remaining knots
does not change. Taking into account that the X-ray intensity can be
defined, as:
\begin{equation}\label{eq2}
  I_X\propto N \frac{L}{4 \pi R^2},
\end{equation}
where $R$ is the physical distance from the quasar, and knowing the
energy density of the cosmic microwave background $W_{CMB}$ and the
ratios of observed intensities at jet knots, $I_A/I_B \approx 2,
I_A/I_C \approx8$, hence, the energy density of quasar emission at
the knot A1 was found: \[W_{ES}^A=8 \cdot W_{CMB} \cdot
\frac{1}{2}.\] The distance $R$ from the knot A1 to 3C273 being
known, the quasar luminosity $L$ has been obtained $L=4 \pi c R^2
W_{ES}^A$. It is consistent with the observed one. Hence, this
assumption is believable.

\section*{Determination of the jet parameters}
\indent \indent Jets enclose the thermal plasma that provides the
main contribution to the jet mass. But due to its low concentration
it cannot be detected directly. The collisions between particles of
the thermal plasma can provoke absorption of the emission. The
absorption coefficient is
\begin{equation}\label{eq3}
\mu \propto \frac{\omega_p^2 \nu}{\omega c},
\end{equation}
where $\omega_p=\sqrt{\frac{4 \pi n e^2}{m}}$ is the the plasma
frequency, $n$ and $e$ are the concentration and the charge of
electrons, $\nu$ is the effective collision frequency.

In this case the energy density of quasar emission is:
\begin{equation}\label{eq4}
W_{ES}=\frac{L}{4 \pi c r^2} \sin^2\theta \exp[-\mu
\frac{(r-r_0)}{\sin \theta}],
\end{equation}
where $r_0$ is the project distance from the quasar to jet
begining, $r$ is the project distance from the quasar,
 $\theta$ is the the angle between jet axis and line of sight.

If the absorption of the quasar emission in the jet does not exist,
then the basic term (\ref{eq1}) is true for $\theta=10^\circ$.
Taking into account that the jet electrons have a Lorenz factor of
$10^3 \div 10^4$, only the photons having the frequency $\omega =
(10^9 \div 10^{13})$Hz before scattering can fall into the Chandra
frequency range $\omega_{Ch}= (10^{17} \div 10^{19})$Hz. The
estimates of the lowest temperature and biggest concentration of the
thermal electrons have been made when the absorption coefficient
tends to zero. We have obtained values for $n\sim 0.1$ cm$^{-3}$ and
$T>10^4$ K.

If $\theta > 10^\circ$, then the basic term (\ref{eq1}) is true when
there is absorption of quasar emission along the jet. The absorption
coefficient should be of the order of $10^{-23}$ cm$^{-1}$. Using
the (\ref{eq1}), (\ref{eq3}) the inequality is:
\[\frac{n^2}{\omega^2 T^{3/2} \sin \theta}>4 \cdot 10^{-23} \ \frac{c^2}{cm^6 K^{3/2}}.\]
We found that either $n>10$ cm$^{-3}$ for $T=(10^3\div 10^4)$ K or
$n\sim 1$ cm$^{-3}$ for $T\sim 10^2$ K. Using (\ref{eq2}) and the
expressions for energy densities of the quasar emission at the
knots A1, B1 and C1
\[W_{ES}^A=\frac{L}{4 \pi c a^2} \sin^2\theta ,\]
\[W_{ES}^B=\frac{L}{4 \pi c (a+b)^2}\sin^2 \theta \exp[-\frac{\mu
b}{\sin\theta}],\]
\[W_{ES}^C=\frac{L}{4 \pi c
(a+2b)^2}\sin^2\theta \exp[-\frac{2 \mu b}{\sin\theta}],\] (where
$a$ is the distance from 3C273 to knot A1, $b$ is the distance from
the knot A1 to knot B1 and the distance between knots B1 and C1(see
Fig.1)), the following inequalities were obtained:
\[\sin^2\theta >W_{CMB}\frac{4 \pi c a^2}{L}(\frac{I_A}{I_B}\frac{N_A}{N_B}),\]
\[\sin^2\theta <W_{CMB}\frac{4 \pi c a^2}{L}(\frac{I_B}{I_A}\frac{N_A}{N_B}).\]
They mean that the energy density of quasar emission appears to
predominate at the knot B1, while the energy density of cosmic
microwave background appears to predominate at the knot C1. From
these inequalities the constraint on $\theta$ value was found:
$18^\circ < \theta <37^\circ$.

\section*{Conclusions}
\indent \indent We have investigated the images of the jet of 3C273
in the different spectral bands to determine mechanism of X-ray
emission from this jet. Having supposed, that the X-ray emission
from the nearest to the quasar knots A1 and B1 are produced by
inverse Compton scattering (IC) of quasar photons. The X-rays from
the jet region with constant X-ray intensity are due to the same
scattering of cosmic microwave background photons. Assuming this the
concentration and temperature of the thermal electrons in the jet of
3C273 were estimated. Obtained values are consistent with
theoretical ones for such objects. Hence, in the such framework we
can estimate not only the angle between the jet axis and the line of
sight, but also the concentration and the temperature of thermal
electrons in the jet of 3C273. We can perform this by comparing jet
images in the X-ray, optical and radio bands.

\section*{Acknowledgements}
\indent \indent We are thankful to Victor M. Kontorovich for
discussions of this work.

\end{document}